\documentstyle[11pt,paspconf,epsf,psfig,twoside]{article}

\def\kmps{km\,s$^{-1}$}
\def\uz{UZ~For}
\def\bl{BL~Hyi}
\def\qq{QQ~Vul}
\def\hu{HU~Aqr}
\def\v13{V1309\,Ori}
\def\degr{\hbox{$^\circ$}}

\begin{document}

\title{Tomography of Polars}


\author{Axel D.~Schwope, Robert Schwarz, Andreas Staude}
\affil{Astrophysikalisches Institut Potsdam, Potsdam, Germany}
\author{Claus Heerlein}
\affil{Institut f\"{u}r Theoretische Physik II, University Erlangen, 
	Germany}
\author{Keith Horne and Danny Steeghs}
\affil{Physics and Astronomy, University of St.~Andrews,
	Scotland}

\begin{abstract}
We are reviewing the power of Doppler tomography for detailed 
investigations of polars (magnetic cataclysmic binaries 
without accretion disk). Using high-resolution spectroscopy in combination
with tomography as analytic tool, the structure and 
extent of the accretion flows and accretion curtains 
in the binaries can be uncovered and the  irradiated and 
non-irradiated parts of the mass-donating secondary stars
can be made visible. In addition we show, how basic system parameters
like the binary inclination and the mass ratio can be addressed by
tomography. 
\end{abstract}


\section{Introduction}
Doppler tomography of disk CVs has become fashionable about a decade 
ago by the pioneering study of Marsh \& Horne (1988). It maps emission
line regions by regarding an observed line profile as projection of the
velocity field along the line of sight. While rotating the binary star 
delivers a set of projections in different directions as the line 
of sight rotates with respect to the line-emitting regions. 
Given time-resolved spectra $f(v, \phi)$ with suitable resolution in radial 
velocity $v$ and binary phase $\phi$, the 2-dimensional constraints from half 
a binary orbit suffice to construct a 2-dimensional map of the system in 
Doppler coordinates $(v_x,v_y)$. Orbital and streaming velocities both 
contribute to the observed Doppler profiles and make Doppler maps 
degenerate in that respect. 
Thus, a unique and straight transformation from Doppler 
space into real space is impossible. 
The interpretation of Doppler maps therefore requires additional 
modelling.

\begin{figure}
\begin{center}
\begin{minipage}{\textwidth}
\psfig{file=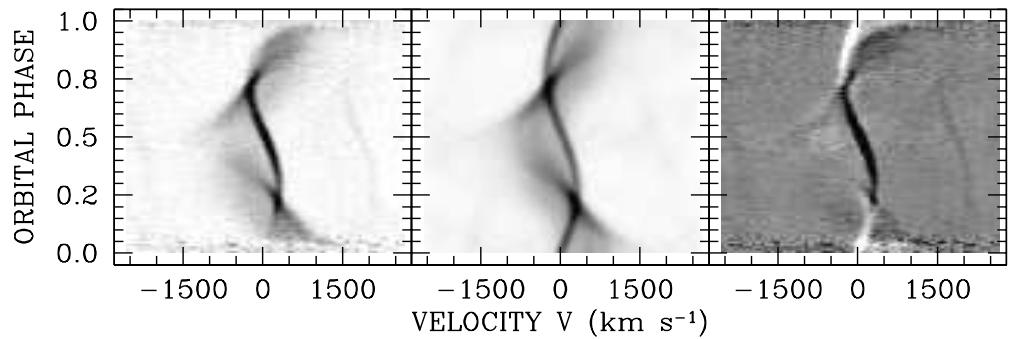,bbllx=74pt,bblly=275pt,bburx=548pt,%
bbury=444pt,width=\textwidth,clip=}
\caption{Trailed spectrogram of the He\,{\sc ii} emission line of \hu\ 
observed in the 1993 high state (adapted from Schwope et al.~1997). The 
observed data has been transformed into Doppler space (Fig.\ref{f:tom_hu})
and projected again into the observers space (middle panel). The residuals
between the two trails shown in the left and middle panels are shown 
right.}
\label{f:opt_thick}
\end{minipage}
\end{center}
\end{figure}

The first application of Doppler tomography to an AM Herculis star (polar), 
VV~Pup, and a candidate AM Her star, GQ~Mus, 
has been described by Diaz \& Steiner (1994a, 1994b).
These attempts, although undertaken with not very high spectral resolution,
have shown the dramatic difference between Doppler maps of 
magnetic and non-magnetic cataclysmic binaries due to the presence of either
accretion disks or accretion streams (e.g.~Kaitchuck et \,al.~1994).
Gas streams have been uncovered also for the first time by means
of Doppler tomography in Algol systems by Richards et\,al.~(1995).
Deep insight into the unique, because bright and eclipsing,  
polar HU~Aqr was possible through high-time 
resolution observations thus clearly revealing the presence of a ballistic
accretion stream, an accretion curtain and a partially shielded secondary 
star. 

In this article we describe recent progress made in this field through 
observations of several polars and the development of theoretical models.

\section{Tomography and polars}
Doppler tomography implicitly assumes that 
emission is completely optically thin and bound to the orbital plane of the 
binary. The latter condition requires that systemic velocities have been
removed from the data. Both prerequisites are violated in polars due
to the presence of optically thick radiating or absorbing 
surfaces and the presence 
of out-of-plane velocitites along magnetic field lines 
and we shortly address the relevance of both effects 
for the interpretation of Doppler maps.

In Fig.~\ref{f:opt_thick} we show trailed spectrograms based on (a)
observations of HU~Aqr in its high accretion state, (b) projection 
of the Doppler map shown in Fig.~\ref{f:tom_hu}, and (c) the difference 
between both. HU~Aqr is an eclipsing polar with a prominent narrow emission
line component (NEL) originating on the irradiated hemisphere of the 
secondary star. Due to the high inclination of the system this component is 
self-eclipsed by the star and visible only when the irradiated hemisphere 
is in view. 
The Doppler map is highly structured 
and allows a detailed view into the binary. It is nevertheless impossible 
by straightforward application of the tomographic inversion process
to reach a good fit to the data 
(small $\chi^2$) due to the self-eclipse of the NEL and the eclipse of the 
other emission components from the accretion stream by the mass-donating 
secondary star. This causes the large systematic residuals of 
Fig.~\ref{f:opt_thick}.
Optical thick emission in the stream itself also 
violates one of the basic assumptions of tomography (but allows mapping of
the observed brightness variation on a given surface, see the contribution 
by Vrielmann \& Schwope in this volume). We mention three recipes which can 
be applied in order to improve on the goodness of fit: (a) use of 
different subsets of the data (in the optical thin case the data from 
half a binary orbit give a full map; one can construct a series of maps,
using for each map half of the data, $\Delta\phi = \phi_{\rm start} + 0.5$,
sequentially shifting the  start phase $\phi_{\rm start}$, and
inspect these series for systematic changes); (b) decomposition of the line 
profiles e.g.~by removing the NEL from the secondary star; (c) mapping of the
lines on predefined surfaces. Rutten \& Dhillon (1994) have established 
with their 'Roche tomography' a first version of this kind.
The more complicated (and much less well-defined) application to accretion
streams and curtains is presently missing.

\begin{figure}
\begin{center}
\begin{minipage}{\textwidth}
\psfig{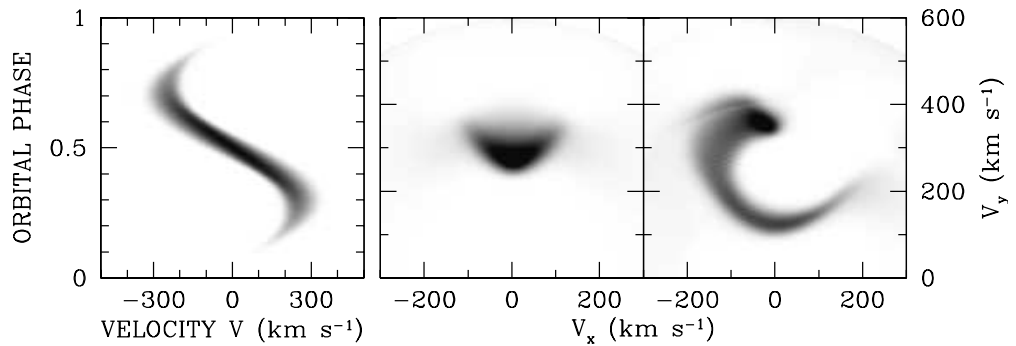}
\caption{Numerical experiments demonstrating the influence of a non-zero 
systemic velocity. {\it (left}) Synthetic trailed spectrogram of an
emission line originating on the Roche lobe of the secondary star 
irradiated by a point source at the white dwarf. {\it (middle)} Reconstructed 
Doppler map with the correct systemic velocity $v_z = 0$\,\kmps.
{\it (right)} Reconstruction with an assumed velocity $v_z = 100$\,\kmps.}
\label{f:sim_gamma}
\end{minipage}
\end{center}
\end{figure}

In order to demonstrate the effect of systemic velocities $\gamma$ 
(or equivalently the effect of out-of-plane velocities) we show in 
Fig.~\ref{f:sim_gamma} 
synthetic line profiles originating on the illuminated hemisphere of the 
secondary star (the NEL) and reconstructions with two 
different values of $\gamma$. The map with the 
correct value $\gamma = 0$\,\kmps\ reconstructs the shape of the Roche lobe 
well (within the limits set by the assumed spectral resolution and
optical thick emission). 
Reconstructions with $v_z = \gamma \neq 0$ 
result in maps with arc-shaped or ring-like 
structures centred on the nominal position
of emission with radius $\sim$$\gamma$. Hence, maps based on spectral data 
with large $\gamma$-velocities or large out-of-plane velocities may be 
significantly blurred. While one can of course correct for non-zero 
$\gamma$-velocities, emission at $v_z \neq 0$ cannot be suppressed or
removed from the data. However, the impact of this kind of emission 
on the maps can become large only for low-inclination systems but these are
generally unfavourable targets for Doppler tomography due to their 
low projected radial velocities in the plane.

\section{Basic binary parameters derived from Doppler maps}

\begin{figure}[bth]
\begin{center}
\begin{minipage}{\textwidth}
\psfig{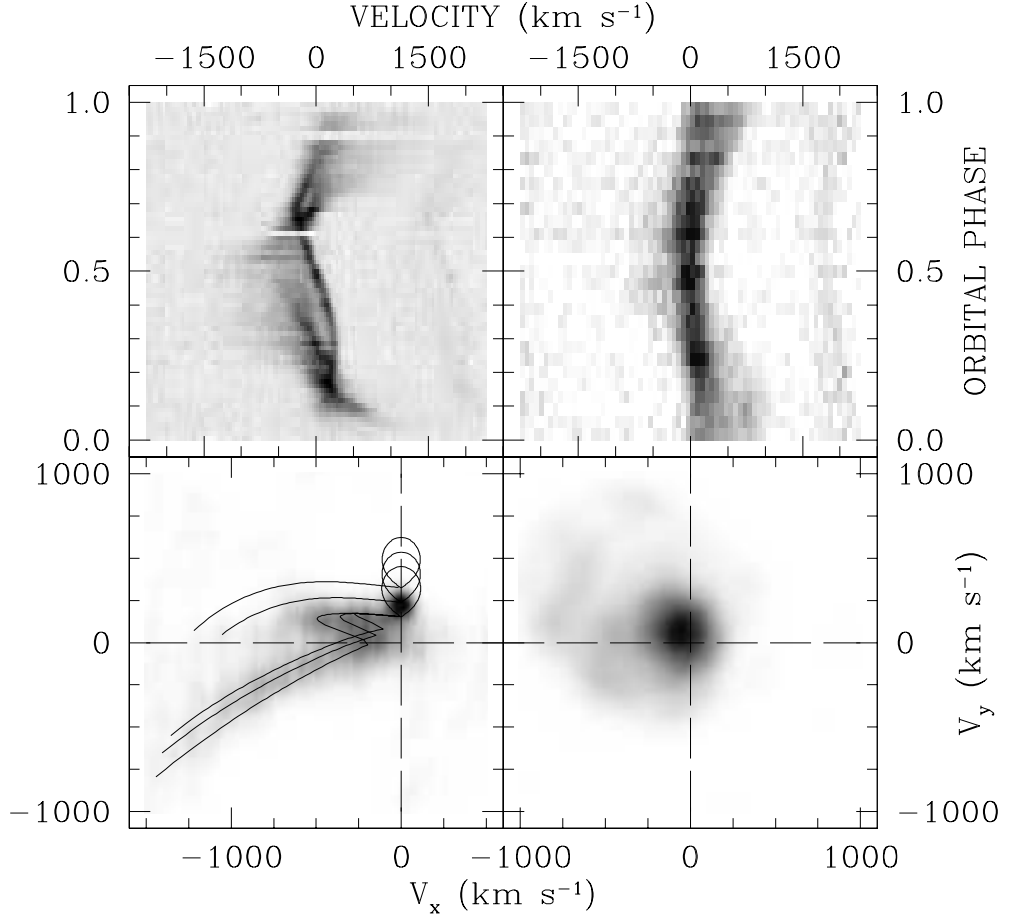}
\caption{Trailed spectrograms (top) and Doppler maps (bottom) of 
the polars \uz\ and \bl\ of the emission line He\,{\sc ii}\,4686\AA. 
The data of \bl\ have only about half the 
spectral resolution than that of \uz. The overlays in the lower 
left panel show Roche lobes of the secondary star and ballistic
trajectories for mass ratios $Q = 8,5,3$ from top to bottom, respectively.}
\label{f:uz_bl}
\end{minipage}
\end{center}
\end{figure}

Fig.~\ref{f:uz_bl} compares trailed spectrograms and Doppler maps of the 
He\,{\sc ii} line at 4686\,\AA\ of the high-inclination, eclipsing system 
UZ~For ($P_{\rm orb} = 126.5$\,min)
and of BL~Hyi with binary period $P_{\rm orb} = 113.6$\,min.
The inclination of the latter system is debated. 

The trailed spectrogram of \uz\ shows several line components displaying 
pronounced radial velocity and intensity variations. In general, this 
behaviour is similar to that seen in the twin system 
\hu\ at almost the same orbital period,
$P_{\rm orb} = 125.0$\,min (Fig.~\ref{f:opt_thick}). The Doppler map
clearly reveals three different structures: (a) the irradiated hemisphere
of the secondary as spot on the axis $v_x = 0$\,\kmps, (b) the ballistic 
part of the accretion stream extending at nearly constant $v_y$ 
from the secondary star down to $v_x \simeq -600 \dots -700$\,\kmps, and (c)
the magnetically dominated part of the stream which starts at about the 
origin and stretches into the lower left quadrant, becoming very 
dim at $(v_x,v_y) \simeq (-1500,-700)$\,\kmps. We refer to the ballistic part 
of the stream also as the horizontal stream, because it is assumed to 
be strictly bound to the orbital plane having no vertical velocity 
component $v_z$.
If one assumes that the centre of
light of the horizontal stream in the Doppler map 
follows a one-particle 
trajectory in the gravitational potential of the binary with mass ratio $Q$
one has a good handle on the value of $Q$. However, 
Schwope, Mantel \& Horne (1997) have shown that in the case of \hu\ the 
horizontal stream in the map and the expected single-particle trajectory
appear apart from each other
(see Fig.~\ref{f:tom_hu}). 
As shown below in Fig.~\ref{f:v13_qq}, \v13\ seems to be 
another counterexample.

Another constraint on the mass ratio comes from the location of the 
bright spot on or near the irradiated surface of the secondary. Again, 
the location of the photocentre can be used to determine the size of the 
Roche lobe and hence the mass ratio $Q$. This approach assumes a certain 
relation between the velocities of the photocentre and the centre of mass and
can be calculated with published irradiation models (Horne \& Schneider 1989;
Beuermann \& Thomas 1990). In \hu\ both methods yielded
values of $Q$ which were mutually exclusive and one had to compromise
on an average value (see Fig.~\ref{f:tom_hu}). 

In \uz\ both methods yield results which are
in agreement with each other.
Roche radii and ballistic trajectories for three different values of $Q$ 
are overlaid on the Doppler map of \uz\ ($Q= M_{\rm wd}/M_2 = 
8, 5, 3$ from top to bottom).
They show best agreement with both features, the ballistic stream and 
the illuminated portion of the 
secondary star for the lowest mass ratio assumed. This picture certainly
excludes the high value of $Q$ (and a high mass of the white dwarf) 
proposed by Beuermann, Thomas and Schwope (1988) based on low-resolution 
spectral observations of the Na-doublet from the secondary 
and is just compatible with 
the range of $Q$'s proposed by Bailey \& Cropper (1991). 
Based on a photometric investigation of the eclipse light curve
the latter authors derived  $Q=5$.

The Doppler map of \uz\ also uncovers the location of the bulk of emission 
from the accretion stream. Assuming a certain orientation and strength 
of the (dipolar) magnetic field the motion of a test particle and its location 
in Doppler coordinates can be followed from the point $L_1$ 
down close to the surface of the white dwarf
(see Schwope et\,al.~1997 for a full description of the model).
With a co-latitude of the magnetic axis 
$\delta =15\degr \equiv 165\degr$ 
and an azimuth $\varphi = 45\degr$ excellent agreement
between the observed and modelled location of the stream can be reached.
The three trajectories shown in Fig.~\ref{f:uz_bl} 
couple onto magnetic field lines $\varphi = 10\degr - 20\degr$ 
prior to eclipse centre.  
These parameters predict a location of the accretion spot at co-latitude 
26\degr ($\equiv 154\degr$), azimuth 31\degr, and, with the azimuth of the 
coupling region of about 15\degr, 
also the occurence of an X-ray absorption dip at phase 0.96. The former value
is in good agreement, the latter two disagree
with the detailed modelling of extended EUVE-observations
by Warren, Sirk \& Vallerga (1995) who observed the dip at phase 0.91
and the spot at an azimuth of 49\degr. This 
leaves three possibilities open:
(1) either our modelling is wrong, or (2) the stream that we are seeing in the 
Doppler maps feeds a secondary accretion spot which 
is different from that seen with EUVE and ROSAT, or (3) our map 
reveals a pronounced re-arrangement of the accretion geometry. Since we
don't believe that (1) applies, a more careful analysis to be performed 
will discern between (2) and (3).

The determination 
of the binary inclination of polars has been the domain 
of polarimetrists for nearly two decades. 
In most cases different investigations
yielded results agreeing with each other within about 10--20\degr. This 
is not true for BL~Hyi where Cropper (1987) and Piirola, Reiz \& Coyne (1987)
derived a high inclination close to eclipse, $i \simeq 70\degr$, whereas
Schwope \& Beuermann (1989) arrived at $i \simeq 30\degr$. 
The trailed spectrogram and the Doppler map shown in Fig.~\ref{f:uz_bl} 
suggest that \bl\ almost certainly has a low orbital inclination. Neither
a NEL from the secondary star with corresponding high radial velocity 
amplitude sticks out nor does the velocities in the stream reach such 
high values as in \uz\ or \hu\ 
nor is a pronounced photometric variability of the line radiation 
recognizable. The Doppler map correspondingly is an unstructured 
mountain of light centred on some location where the stream and the 
secondary star are expected. The radial velocity curve of the emission 
lines observed in BL Hyi reveals a velocity $v_z \simeq 
120$\,km s$^{-1}$ if the lines in individual spectra are 
fitted by single Gaussians. Hence, the
blurring effect described in the previous section additionally 
smears the map. 
 
\begin{figure}[t]
\begin{center}
\begin{minipage}{\textwidth}
\psfig{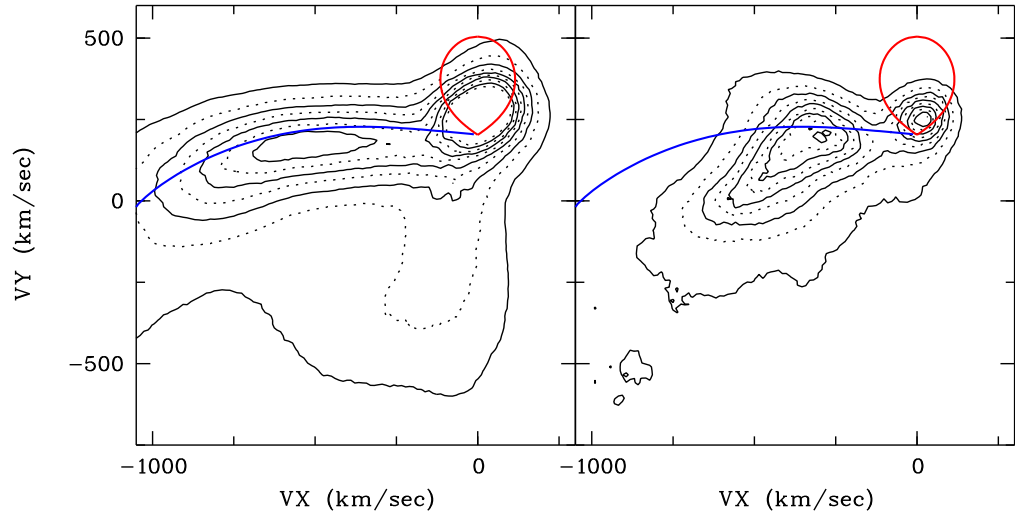}
\caption{Doppler maps of the eclipsing polar \hu\ in the light of 
the He\,{\sc ii}\,4686 emission line observed in the 1993 
high (left) and the 1996 reduced state of accretion (right).
The overlays indicate the size of the Roche lobe of the secondary 
and the ballistic trajectory for a mass ratio $Q=4$.
}
\label{f:tom_hu}
\end{minipage}
\end{center}
\end{figure}

\begin{figure}[t]
\begin{center}
\begin{minipage}{\textwidth}
\psfig{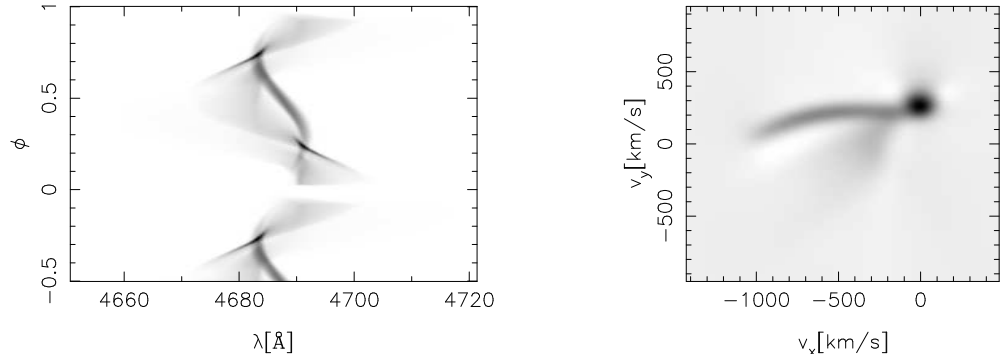}
\caption{Simulated trailed spectrogram and Doppler map for \hu\
in its high accretion state
(Heerlein et al.~1998).
}
\label{f:mod_hu}
\end{minipage}
\end{center}
\end{figure}

\section{HU~Aqr: raising and lowering the accretion curtain}
There is only one system which was observed with sufficient high spectral
and time resolution in different accretion states allowing Doppler 
tomography, \hu\ (Fig.~\ref{f:tom_hu}, Schwope et\,al.~1997, 1998).

At both occasions the tomograms show an asymmetrically irradiated secondary 
star. Recombination radiation from the secondary is always
centred at $v_x > 0$\,\kmps\ which means that some kind of accretion curtain 
was always shielding the leading side of the star. The major difference 
between both maps concerns the emission from the accretion stream. In the high 
accretion state the stream penetrates the magnetosphere along the 
ballistic trajectory much deeper than at reduced accretion which seems 
to be obvious but nevertheless is directly confirmed by imaging of the 
stream for the first time. When coupling onto 
field lines the velocity vector of a plasma cell is assumed to be changing 
over a very short distance in real space. This leads to a jump of the 
stream in Doppler coordinates $(v_x,v_y)$. The stream 'turns sharper around the
corner' in the high state which results in a larger jump towards the origin
than in the low state. Hence the appearance of both parts of the stream 
in the Doppler map is affected by the change of the accretion rate. 
The one-particle ballistic trajectory 
plotted in Fig.~\ref{f:tom_hu} was computed for $Q = 4$. It does not coincide
with the center of light of the observed ballistic stream but it represents 
the best compromise between estimates of $Q$ based on the location of the 
ballistic stream and the NEL in the map. The reason for the displacement 
is not understood.

A further step towards a detailed understanding of the trailed spectra 
of polars in general but of \hu\ in its high accretion state in particular has
been  undertaken recently by Heerlein, Horne \& Schwope (1998). They developped
a still quite simple stripping model for the accretion stream in three 
dimensions and considered radiation reprocessing in optically thick 
approximation from the ballistic stream and the accretion curtain 
connected to it. Irradiation of the companion star was included in their 
code but without shielding of the secondary by the curtain. 
The length of the ballistic stream, the size of and the velocities
in the curtain are depending on a few basic system parameters, mass ratio, 
accretion rate, orientation and strength of the magnetic field. By variation 
of either parameter the trailed spectra change, the predicted trailed 
spectrogram for the least-squares solution achieved by Heerlein et\,al.~is 
reproduced in Fig.~\ref{f:mod_hu} together with the corresponding Doppler 
map. The images shown there should be compared with Figs.~\ref{f:opt_thick}
and \ref{f:tom_hu}. The model reflects well the main observed features, 
it can, however, not explain the displacement of the horizontal stream.
It also predicts a significantly better focussed ballistic stream than 
the observations show (the models were using the same spectral 
resolution as the data). A solution for the discrepancies lies probably 
in the details of the fragmentation of the stream, its more complex 
inner structure and a more complicated magnetic field topology. 

Finally we would like to mention that in the low accretion state a 
significant amount of emission is seen in the upper left quadrant above 
the ballistic trajectory which is also not compatible with our current 
understanding of the streams in polars and suggestive of a complex 
flow pattern in the magnetosphere.

\begin{figure}
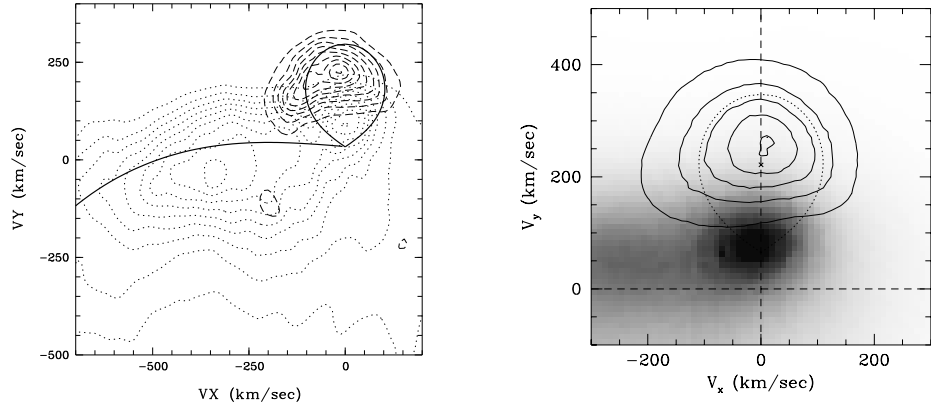

\begin{minipage}{55mm}
\psfig{file=v13_ps,%
bbllx=57pt,bblly=115pt,bburx=541pt,bbury=613pt,angle=-90,clip=,width=55mm}
\end{minipage}
\hspace*{1cm}
\begin{minipage}{55mm}
\psfig{file=qq_ps,%
bbllx=83pt,bblly=232pt,bburx=511pt,bbury=638pt,clip=,width=55mm}
\end{minipage}
\caption{Combined Doppler maps of He\,{\sc ii} emission lines and 
Na\,{\sc i} absorption lines of the long-period eclipsing polar
\v13\ (left) and of \qq\ (right). Overplotted are the Roche lobes
of the secondary stars for mass ratios $Q=1.75$ and $Q=1.71$, respectively.
The maps of the Na-lines are shown as contour plots (dashed contours left,
solid contours right), these lines
are clearly visible only on the non-irradiated side of the companion stars.
}
\label{f:v13_qq}
\end{figure}

\section{Irradiated companion stars}
Fig.~\ref{f:v13_qq} compares the He\,{\sc ii} emission and Na\,{\sc i} 
absorption lines of \v13\ and \qq. In these long-period systems the 
secondary is big enough not to become outshone by the stream and the accretion 
spot even in their high accretion states. Tha Na\,{\sc i} lines are originating
in the photospheres of the mass-donating late-type  stars. The
He\,{\sc ii} emission from these secondaries originates from a 
quasi-chromosphere which is formed above the photosphere on response
to the intensive EUV/X-ray irradiation from the hot spot on the white dwarf.
The luminosity of the irradiating source is higher than that of the 
exposed star and apart from possible partial shielding by an accretion curtain
these stars are fully exposed.

To our knowledge there are no model calculations available in the literature
which may be applied directly to this situation. The situation is similar
to the HZ\,Her/Her\,X-1 system were several models with different level
of sophistication has been developped (e.g., London et\,al.~1981) but the type
of the irradiated  star and of the irradiating source are clearly 
different here.
Brett \& Smith (1993) have developped a first model applicable to CVs 
irradiation by a moderately hot (17000\,K) white dwarf in LTE. They took 
in favour of their approach King's (1989) argument that the ionizing radiation 
cannot reach the photosphere. Their model showed already that the temperature
structure and the emitted spectrum of the irradiated star are changed 
markedly.

Our data clearly support this view Na\,{\sc i} emission in \qq\ and \v13\ 
is strongly supressed if visible at all on the irradiated hemisphere. 
This seems to be the common picture in polars, other examples are AM\,Her 
(Davey \& Smith 1992) and \hu\ (unpublished data from the authors) and tells
us that all attempts to determine the stellar masses 
(in particular $M_{\rm wd}$) by 
radial velocity measurements of the secondary stars 
have to pay attention to this distortion. 
The detection of the irradiation distortion requires 
high resolution spectroscopy (dispersion and time) of faint absorption 
lines in a spectral region which is highly contaminated by night sky 
emission and atmospheric absorption lines and is, therefore, an
ambitious observational task.

\section{Conclusions and outlook}
Doppler tomography of polars has become possible in recent years through 
the mounting of sensitive low-noise detectors in high-resolution spectrographs
at 4m-class telescopes. The angular resolution thus achieved is of the order 
of {\it micro arcsec}. Images of accretion streams, accretion curtains
and the two sides of X-ray illuminated M-stars could be gained.
Interpretation of these Doppler images has just started and opens the 
field for a physical understanding and modelling of the accretion plasma
between the two stars. We mention a few topics for future work in that 
direction: (1) The width of the ballistic plasma streams and their 
(occasionally) displaced location with respect to a one-particle 
trajectory are not matched by purely kinematical considerations.
Turbulence and MHD effects obviously have to be taken into account 
(although this will be difficult to achieve in practice due to the 
complexity of the problem). (2) When modelling the emission from the 
stream/curtain region in this paper we have made use mainly of the 
velocity information. Future work should address also the line 
intensities (in different line components, in different species) 
and try to derive the temperature/ionization structure by application 
of photoionization models. (3) The secondary stars provide a natural laboratory
for the application of so-called next generation stellar models (NGSM, e.g.,
Hauschildt, Baron \& Allard 1997) taking into account irradiation 
by the hot spot on the white dwarf. Again, this is not an easy and 
straightforward task due to NLTE and the anisotropy but models and
observations are approaching a state where this kind of research becomes
feasable.

Finally, our view into these type of binaries can be sharpened by 
utilization of future 8m telescopes, e.g.~the VLT UT2 with UVES. We will
be able then to look e.g.~for starspots on the secondary stars by means
of Doppler imaging and thus be able to address the question if giant
starspots might be responsable for the low states in cataclysmic binaries.

\acknowledgments
This work has been supported through DLR grants 50 OR 9403 5 and 50 OR 9706 8
and by the DFG grant Schw 536/8-1.

\end{document}